\documentclass[12pt]{article}
\pdfoutput=1
\usepackage{subfigure}
\usepackage{amssymb,amsmath}
\usepackage{graphicx}
\usepackage{color}
\usepackage[colorlinks=true
,urlcolor=blue
,citecolor=blue
,linkcolor=blue
,pagecolor=blue
,linktocpage=true
,pdfproducer=medialab
]{hyperref}
\usepackage[a4paper]{geometry}

\makeatletter \renewcommand{\@dotsep}{10000} \makeatother

\def\mmess{M_{\rm mess}}
\mathchardef\mhyphen="2D

\newcommand{\beq}{\begin{equation}}
\newcommand{\eeq}{\end{equation}}
\newcommand{\bea}{\begin{eqnarray}}
\newcommand{\eea}{\end{eqnarray}}

\begin{document}

\begin{titlepage}
\pagestyle{empty}

\vspace*{0.2in}
\begin{center}
{\Large\bf    Reconciling Muon $g-2$, 125 GeV Higgs and Dark Matter in Gauge Mediation Models
  }\\
\vspace{1cm}
{\large Ilia Gogoladze$^\ast$\footnote{E-mail: ilia@bartol.udel.edu;\\
\hspace*{0.5cm} On leave of absence from Andronikashvili Institute
of Physics,  Tbilisi, Georgia.},
Qaisar Shafi$^\ast$\footnote{E-mail: shafi@bartol.udel.edu} and
Cem Salih $\ddot{\rm U}$n$^{\diamond, \dagger}\hspace{0.05cm}$\footnote{E-mail: cemsalihun@uludag.edu.tr}}
\vspace{0.5cm}

{\it
$^\ast$Bartol Research Institute, Department of Physics and Astronomy,\\
University of Delaware, Newark, DE 19716, USA \\
$^\diamond$Department of Physics, Uluda\~{g} University, TR16059 Bursa, Turkey \\
$^\dagger$ Center of Fundamental Physics, Zewail City of Science and Technology, \\
6 October City, Cairo 12588, Egypt}

\end{center}

\vspace{0.5cm}
\begin{abstract}
\noindent

 We present a class of models in the framework of gauge mediation supersymmetry breaking where the standard model is supplemented by additional U(1) symmetry which acts only on the third generation fermions. The messenger fields carry non-trivial U(1) charge and are vector-like particles under this symmetry. This leads to additional contribution to the soft supersymmetry breaking mass terms for the third generation squarks and sleptons. In this framework we show that the muon $g-2$ anomaly, the observed 125 GeV Higgs boson mass and the detected relic dark matter abundance (gravitino in our case) can be simultaneously accommodated. The resolution of the muon $g-2$  anomaly, in particular, yields the result that the first two generation squark masses, as well the gluino mass, should be  $\lesssim 2.5$ TeV, which will be tested at LHC14.

\end{abstract}

\end{titlepage}

\section{Introduction}

The discovery of the Higgs boson  with a mass $ \sim 125$ GeV \cite{:2012gk,:2012gu} has important consequences in general for low scale supersymmetry (SUSY). In the minimal supersymmetric standard model (MSSM) it requires either a large, ${\cal O} (\mathrm{few}-10)$ TeV,  stop quark mass, or alternatively a relatively large soft supersymmetry  breaking (SSB) trilinear $A_t$-term, along with a stop  quark mass of around a TeV  \cite{Heinemeyer:2011aa}. The constraints related to the Higgs boson are  particularly stringent regarding the  sparticle spectrum in the  gauge mediated SUSY breaking (GMSB) scenario \cite{Ajaib:2012vc}.  In both the minimal (mGMSB) \cite{Dine:1993yw} and more general \cite{Meade:2008wd} GMSB scenarios, the trilinear SSB $A-$terms are relatively small at the messenger scale and hence, accommodating the light CP-even Higgs boson mass of 125~GeV requires a stop mass in the  multi-TeV range \cite{Ajaib:2012vc}.  This, in turn, pushes the remaining sparticle mass spectrum including sleptons to values above the TeV scale or so \cite{Ajaib:2012vc}. Imposing $t-b-\tau$ (or $b-\tau$) Yukawa coupling unification condition at $M_{{\rm GUT}}$ in the GMSB scenario makes the sparticle spectrum even heavier than this. It was shown in Ref \cite{Gogoladze:2015tfa} that this latter scenario predicts that all colored sparticle masses are above 3 TeV, which would be hard to test at LHC 14 but  may be accessible at some future colliders such as HE-LHC 33 TeV or a 100 TeV collider.

The Standard Model (SM) prediction for the anomalous magnetic moment of the muon, $a_{\mu}=(g-2)_{\mu}/2$ (muon $g-2$) \cite{Hagiwara:2011af}, has a  $(2-3)\, \sigma$ discrepancy with the experimental results \cite{BNL}:
\begin{eqnarray}
\Delta a_{\mu}\equiv a_{\mu}({\rm exp})-a_{\mu}({\rm SM})= (28.6 \pm 8.0) \times 10^{-10}\,.
\end{eqnarray}
If supersymmetry is to offer a  solution to this discrepancy, the   smuon and gaugino (bino or wino)   should  be
relatively light,  ${\cal O}(100)$ GeV. Thus, it is hard to simultaneously explain
the observed Higgs boson mass and resolve the muon $g-2$ anomaly with mGMSB (mGMSB).
There have been several recent attempts to reconcile this (presumed) tension between
muon $g-2$ and 125 GeV Higgs mass in the mGMSB framework. For instance, adding low mass vector-like particles \cite{Gogoladze:2009bd} can provide significant contributions to the Higgs boson mass \cite{Endo:2011mc}. Another suggestion is to introduce  matter-messenger mixing \cite{Evans:2012hg},
but this  can potentially reintroduce  the SUSY
flavor problem in mGMSB, which is undesirable. Ref. \cite{Evans:2012hg} assumes a perfect alignment of appropriate parameters in order to avoid unwanted flavor violating processes. Alternatively, a specific flavor symmetry is employed to overcome this challenge \cite{Albaid:2012qk}.
It was shown in Ref. \cite{Bhattacharyya:2013xma} that by enlarging the messenger sector with a weak triplet and a color octet. one can resolve the muon $g-2$ anomaly and obtain a 125 GeV Higgs boson in the spectrum. In order to satisfy the cosmological and accelerators constraints in this model,  R-parity violating interactions are introduced.

Recently, Ref. \cite{Babu:2014sga} proposed a class of supersymmetric models in the framework of gravity mediated supersymmetry breaking \cite{Chamseddine:1982jx},  in which symmetry considerations alone dictate the form of the soft SUSY breaking (SSB) mass terms for the sfermions \cite{Babu:2014sga}.  It is called flavor symmetry based MSSM (sMSSM), and in this framework the first two family sfermion masses have degenerate masses at $M_{{\rm GUT}}$, while the SSB mass term for third family sfermions is different. It was shown in \cite{Baer:2004xx} that constraints from flavor changing neutral current (FCNC) processes, for the case when third
generation sfermion masses are split from masses of the first and second generations,
are very mild and easily satisfied. This approach therefore allows for significantly lighter first two family sfermions, while keeping the third generation sfermions relatively heavy. The first two families and gauginos (bino and wino) play an important role in the resolution of the muon $g-2$ problem.
The authors in Ref. \cite{Baer:2004xx,Ibe:2013oha} assume MSSM Higgs fields ($H_u$ and $H_d$)  with a common SSB mass either with the first two generation sfermions or with the third generation squarks and sleptons. In these  approaches it was shown \cite{Ibe:2013oha} that  one can easily  accommodate the resolution of muon $g-2$ problem with the 125 GeV Higgs boson mass, but it is difficult to obtain  neutralino dark matter with the correct relic abundance. Ref. \cite{Babu:2014lwa} proposed to consider  SSB mass terms for $H_u$ and $H_d$ independent from both the first two generation and third generation sfermion SSB mass terms. As a consequence, it is possible to simultaneously have 125 GeV Higgs boson, solve the muon $g-2$ anomaly and accommodate the correct relic abundance using neutralino- slepton coannihilation. Moreover,  this framework allows one to implement $t$-$b$-$\tau$ Yukawa unification \cite{Ajaib:2014ana}.

In this paper we propose two models which allow us to realize the sMSSM structure in the GMSB framework, with relatively light and nearly degenerate squark and slepton masses of the first two families. The third generation sfermions are heavy with masses $\sim 10$ TeV. One simple way to realize this  scenario  within GMSB  is to introduce a flavor dependent $U(1)$ gauge symmetry, and we present two distinct scenarios  here. We first consider $U(1)_{1}$, under which all the left-handed fermions  from the third generation have charge ($+\eta$), while  the third family right-handed fermions (including  right-handed tau neutrino) have charge ($-\eta$). We also assume that the messenger fields (for instance $(5+\bar5)$) transform non-trivially  under  $U(1)_1$.  Thus, the messenger fields interact with the third generation sfermions  via $U(1)_1$ gauge interaction and provide additional contributions to their SSB mass terms.  This new contribution makes the third generation sfermions adequately heavy, while the first two family sfermions remain light and degenerate in mass. Note that we also increase the contribution to the trilinear SSB breaking terms for third generation sfermions through the $U(1)_1$ gaugino loop, but this additinoal contribution turns out to be negligible. The model allows mixing between the first two generation fermions with the third generation via a non-renormalizable coupling. However, the desired mixing between the generations which yields the correct Cabibbo–-Kobayashi-–Maskawa  (CKM) mixing  matrix is realized after spontaneous breaking of $U(1)_{1}$.

In Model II a $U(1)_2$  symmetry acts only on the third generation fermions. In this case the left-handed top and bottom fields have  a $U(1)_2$ charge of $+\eta$, while the left-handed tau lepton charge is $-3\eta$. The right-handed top and bottom fields have charge $-\eta$, and the right-handed tau and tau neutrino have charge $+3\eta$. Clearly,  in Model II the $U(1)_2$ charge assignment resembles the well-known $U(1)_{B-L}$ assignments, except that here $U(1)_{2}$ is a flavor symmetry which acts only on the third generation fermions.

This paper is organized as follows: We describe Models I and II in Section \ref{sec:model} and in Section \ref{sec:scan} we summarize the scanning procedure and the experimental constraints we employ. In Section \ref{sec:results} we present our results focusing on the resolution of muon $g-2$ and accommodating the 125 GeV Higgs boson mass and relic dark matter abundance. We also provide two tables of benchmark points in this section, which exemplify our findings. Our conclusions are presented in Section \ref{sec:conclusion}.

\section{Essential Features of the Model}
\label{sec:model}

Supersymmetry breaking in a typical GMSB scenario takes place in a hidden sector, and this effect is communicated to the visible sector via messenger fields. The latter interact with the visible sector via the SM gauge interactions, and induce the SSB terms in the MSSM through loops.
In order to preserve perturpative gauge coupling unification, the minimal GMSB   scenario  can include $N_5$  $(5+\overline{5})$  ($N_{5}=1, ... , 5$) $SU(5)$ multiplets or one $(10+\overline{10})$ pair, or $10+\overline{10}+5 +\overline{5}$, or  one pair of $15+\overline{15}$. For simplicity, we only consider  the case with $N_5$ $(5+\overline 5)$ vectorlike multiplets. Also it is known \cite{Ajaib:2012vc} that the sparticle spectrum does not change drastically for $N_5>1$  compared to the $N_5=1$ case. The   $(5+\overline 5)$  includes $SU(2)_L$ doublets $(L +\bar{L})$, and  $SU(3)_c$ triplets $(q_{c}+\bar{q}_{c})$. Note that the   $(5+\overline 5)$ fields carry additional $U(1)_{1}$ (or $U(1)_{2}$) charges ($\eta +  (-\eta)$).
In order to incorporate SUSY breaking in the messenger sector,  the fields in ($5+\overline 5$) multiplets are coupled, say, with the hidden sector gauge singlet chiral field $S$:
\begin{equation}
W \supset y_1 S\, \ell \, \bar{\ell}+y_2 S\,  q \, \bar{q},
\end{equation}
where $W$ denotes the appropriate superpotential. Assuming non-zero vacuum expectation values (VEVs) for the scalar and $F$ components of $S$, namely $S=\langle S \rangle+\theta^2 \langle F \rangle$, the mass spectrum of the messenger fields is as follows:
\begin{equation}
  m_{b} = M \sqrt{ 1 \pm {\Lambda \over M} }, \ \
m_{f} = M.
\end{equation}
  Here $m_{b}$ and $m_{f}$ denote the masses of the bosonic and fermionic components  of the  appropriate messenger superfield, $M = y \langle S \rangle$ and
  $\Lambda = \langle F \rangle / \langle S \rangle$.
The dimensionless parameter $\Lambda /M$ determines the  mass splitting between the scalars and fermions in the messenger multiplets.
This breaking is transmitted to the MSSM particles via loop corrections.

 At the messenger scale we have the gauge symmetry $SU(3)_c\times SU(2)_L  \times U(1)_Y \times U(1)_1 $ (or $U(1)_2$), where the $U(1)$ charge assignments are listed in Table \ref{table 1}. The  gaugino masses are generated at  1-loop level, and assuming $\langle F \rangle \ll \langle S \rangle^2$,  are given by
\begin{equation}
M_{i} = N_5\, \Lambda \, \frac{\alpha_i}{4\pi}  ,
\label{gauginomass}
\end{equation}
 where $i = 1,\, 2,\, 3,\, 4$  stand for the  $SU(3)_c$,  $SU(2)_L$,  $U(1)_Y$ and $U(1)_1$ (or $U(1)_2$)  sectors, respectively.
The MSSM scalar masses  are induced at  two loop level:
\begin{equation}
m^2(M) = 2\,  N_5\,  \Lambda^2 \,  \sum_{i=1}^3 \, C_i
  \left( \alpha_i \over 4 \pi \right)^2,
\label{scalarmass}
\end{equation}
where $C_1 = 4/3$,   $C_2 = 3/4$,
$C_3 = (3/5) (Y/2)^2$ and $C_4 = \eta^2$. Here $Y$ and $\eta$ denote the hypercharge and $U(1)_1$ (or $U(1)_2$) charges respectively.

\begin{table}[h!]
\centering
\scalebox{0.89}{
\begin{tabular}{|l|l|c|c|c|c|c|c|}
\hline

  &      &$ Q_3$ & $U^c_3 $& $d^c_3$&$L_3$  & $e^c_3$   & $\nu_{\tau_{3}}$  \\

\hline
          &$U(1)_Y$      &$ 1/6$ & $-2/3$& $1/3$ & $-1/2$  & $1$   & $0$  \\
\hline
Model I    & $U(1)_1$   & $-\eta$  & $\eta$ & $\eta$ &  $ -\eta$ &  $ \eta$ &  $ \eta$ \\
\hline
Model II    & $U(1)_2$   & $\eta$  & $-\eta $ &$ -\eta$ & $-3\eta$ &  $ 3\eta $ &  $ 3\eta$ \\
\hline
\end{tabular}}
\caption{ $U(1)$ charge assignments for the third generation fermions. Here we use the standard notation for the SM fermions. $U(1)_Y$ stands for the  hypercharge and $U(1)_1$ (and $U(1)_2$) are additional gauge symmetries that we introduce for Model I (and Model II).}
\label{table 1}
\end{table}

As seen from Eq.\,(\ref{scalarmass}), the extra  contribution to the sfermion  SSB masses from $U(1)_1$ (or $U(1)_2$)
gauge mediation
is proportional to $C_4 = \eta^2$ and $\alpha_4$. A suitable choice of these parameters can make the contribution from $U(1)_1$ (or $U(1)_2$)
gauge mediation independent from what we have in mGMSB. Of course, this contribution cannot be completely arbitrary, but we assume it to be ten times or so larger compared to the contribution from  $SU(3)_c\times SU(2)_L  \times U(1)_Y$.
Note that the $U(1)_{1}$ (or $U(1)_{2}$) contribution, denoted as $D$, in principle, can be quite independent from the $SU(3)_c\times SU(2)_L  \times U(1)_Y$ sector if we introduce messenger fields with only $U(1)_1$ (or $U(1)_2$) charges, and which are singlets under the SM gauge symmetry. In this case we do not need to worry about $C_4 = \eta^2$ and $\alpha_4$ values at all. We can just use proper values of $\Lambda^{\prime}$ in Eqs. (5) and  (6). Since the origin of new contributions to the third generation sfermions can not be tested at the experiment we do not specify the model and added just a new parameter D at the messenger scale which measures the new contribution to the third family.

The summary of model building part of the scenario.  New contribution to the third generation sfermions are obtained
by extending the gauge symmetry by an additional U(1) symmetry compared to the canonical MSSM  gauge mediation scenario.
Note that based on the charge assignment presented in Table 1  the Model I has a chiral anomaly under the  new  $U(1)_1$ symmetry. This anomaly  can be canceled by introducing  set of new particles. For simplicity, we assume that the $U(1)_1$ symmetry is broken not far below the messenger scale and these extra particles  introduced for just cancellation of the chiral anomaly, and hence they obtain masses around  the $U(1)_1$ symmetry breaking scale. So below the messenger scale RGE evaluation is exactly the same as what we have in MSSM. Existence of a new U(1) symmetry in our model changes only the boundary conditions on the SSB mass terms for the third generation.  Since there are a number of ways to cancel $U(1)_1$ anomalies and the new particles are superheavy we do not specify them here.

On the other hand the situation is much simpler for Model II case.  The $U(1)_2$ charge assignments  resemble  $U(1)_{B-L}$ charges  (see Table 1) except that the former is a flavor symmetry which acts only on the third generation fermions. In this context, Model II is very minimal in terms of additional particles in comparison to the canonical MSSM gauge mediation scenario. We just need to introduce the field which can break spontaneously  $U(1)_{B-L}$ symmetry. Again here for simplicity, we assume that  $U(1)_{B-L}$ symmetry is broken just below the messenger scale; therefore, the RGE is the same as which we have in the MSSM case.

 The A-terms in GMSB models vanish  at the messenger scale (except when the MSSM and messenger fields mix \cite{Evans:2015swa}, which we do not consider in this study). They are generated from the RGE running below the messenger scale. The sparticle spectrum in our model(s)  is therefore completely specified by the following parameters defined at the messenger scale:
\begin{equation}
M_{\mathrm{mess}}, \, \Lambda,  \, \mathrm{tan}\beta,  \, sign({\mu}), \,  N_5,\,   c_{\rm grav}, \, D.
\label{mgmsb-params}
\end{equation}
Here $M_{\mathrm{mess}}\equiv M$  and $\Lambda$ are the messenger and SSB mass scales defined above , and ${\rm tan\beta}$ is the
ratio of the VEVs of the two MSSM Higgs
doublets. The magnitude of $\mu$, but not its sign, is determined by the
radiative electroweak breaking (REWSB) condition. The parameter $c_{\rm grav} (\geq 1)$ affects the mass of the gravitino and we set it equal to unity from now on. For simplicity, we consider the case $N_5=1$. Changing the  value of $N_5$ does not significantly alter the sparticle spectrum \cite{Ajaib:2012vc}.

 In summary,  the MSSM sfermion  masses have the following expression:

\begin{equation}
 m_{\phi_{1,2}}^{2} =(m_{\phi_{1,2}}^{2})_{{\rm GMSB}},\hspace{1.0cm} m_{\phi_{3}}^{2}=(m_{\phi_{3}}^{2})_{{\rm GMSB}}+e_{\eta}^{2}D^{2}
 \label{sGMSB_masses}
 \end{equation}
where $\phi_{1,2}$ and $\phi_{3}$ denote the sfermions of the first two families and the third family respectively, and $e_{\eta}$ stands for the sparticle charges under $U(1)_{1}$ (and $U(1)_{2}$). We can normalize the sparticle charges by setting  $\eta=1$.

\section{Scanning Procedure and Experimental Constraints}
\label{sec:scan}

We employ the ISAJET 7.84 package \cite{ISAJET} to perform random scans over the fundamental parameter space. In this package, the weak scale values of gauge and third generation Yukawa couplings are evolved to $M_{{\rm mess}}$ via the MSSM RGEs in the $\overline{DR}$ regularization scheme.The various boundary conditions are imposed at $M_{{\rm mess}}$ and all the SSB parameters, along with the gauge and Yukawa couplings, are evolved back to the weak scale $M_{{\rm Z}}$. The threshold corrections \cite{Pierce:1996zz} are added to the Yukawa couplings at the common scale $M_{{\rm SUSY}} =\sqrt{m_{\tilde{t}_{L}}m_{\tilde{t}_{R}}}$, where $m_{\tilde{t}_{L}}$ and $m_{\tilde{t}_{R}}$ are the soft masses of the third generation left and right-handed stop quarks respectively. The entire parameter set is iteratively run between $M_{{\rm Z}}$ and $M_{{\rm mess}}$, using the full 2-loop RGEs, until a stable solution is obtained.	 To better account for leading-log corrections, one-loop step beta functions  are adopted for the gauge and Yukawa couplings, and the SSB parameters $m_{i}$ are extracted from RGEs at multiple scales: at the scale of its own mass, $m_{i}=m_{i}(m_{i})$, for unmixed sparticles, and the common scale $M_{{\rm SUSY}}$ for the mixed ones. The RGE-improved one-loop effective potential is minimized at $M_{{\rm SUSY}}$, which effectively accounts for the leading 2-loop corrections. Full 1-loop radiative corrections \cite{Pierce:1996zz} are incorporated for all sparticle masses.

We have performed random scans over the model parameters given in Eq.(\ref{mgmsb-params}) in the following range:
\begin{eqnarray}
10^{4}\,{\rm GeV}\leq & \Lambda & \leq 10^{7}\,{\rm GeV}, \nonumber \\
\Lambda ~{\rm GeV}\leq & \mmess &\leq 10^{16}\, {\rm GeV}, \nonumber \\
0 \leq & D & \leq 30 \,{\rm TeV}  \label{parameterRange} \\
2 \leq & \tan\beta & \leq 60,  \nonumber\\
N_{5}=1, & \mu > 0 & m_{t}=173.3 \,{\rm GeV}. \nonumber
\end{eqnarray}
$N_{5}$ can be varied from 1 to 4, but the sparticle spectrum does not change significantly\cite{Ajaib:2012vc}, thus we set $N_{5}=1$ for entire scan of the parameter space. Regarding the MSSM parameter $\mu$, its magnitude but not its sign is determined by the radiative electroweak symmetry breaking (REWSB). In our model, we set $sgn(\mu)=1$. Finally, we employ the current central value for the top quark mass, $m_{t}=173.3 $ GeV. Our results are not too sensitive to one or two sigma variation of $m_{t}$ \cite{Gogoladze:2011db}.

In scanning the parameter space, we employ the Metropolis-Hastings algorithm as described in Ref. \cite{Belanger:2009ti}. The data points collected all satisfy the requirement of REWSB. After collecting the data, we impose the mass bounds on the particles \cite{pdg} and use the IsaTools package \cite{isatools} to implement the various phenomenological constraints. We successively apply mass bounds including the Higgs \cite{:2012gk,:2012gu} and gluino masses \cite{gluinoLHC}. In addition, we apply the mass bounds on selectrom/smuon as $m_{\tilde{\mu}}\geq 245$ GeV  if the smuon is NLSP  \cite{Aad:2014vma} and decays inside the LHC detector. But if  the selectrom/smuon  are long live enough in order to decay outside of detector then $m_{\tilde{\mu}}\geq 400$ GeV \cite{Calibbi:2014pza}. Beside the mass bounds, we also apply the constraints from the rare decay processes $B_{s}\rightarrow \mu^{+}\mu^{-}$ \cite{BsMuMu}, $b\rightarrow s\gamma$ \cite{Amhis:2012bh} and $B_{u}\rightarrow\tau\nu_{\tau}$ \cite{Asner:2010qj}. The constraints are summarized below:
\begin{eqnarray}
   123\, {\rm GeV} \leq m_h \leq127 \,{\rm GeV} &
\nonumber \\
 m_{\tilde{g}} ~({\rm and }~ m_{\tilde{Q}}) \geq1.5 \,{\rm TeV} & \nonumber \\
\hspace{-6.0cm}m_{\tilde{\mu}_{R}} \geq 245 \, (400)\, {\rm GeV} & ({\rm if \,\, smuon \,\, is \,\, NLSP)}  \nonumber \\
0.8\times 10^{-9} \leq{\rm BR}(B_s \rightarrow \mu^+ \mu^-)
  \leq 6.2 \times10^{-9} \;(2\sigma) &
 \\
2.99 \times 10^{-4} \leq
  {\rm BR}(b \rightarrow s \gamma)
  \leq 3.87 \times 10^{-4} \; (2\sigma) &
\nonumber \\
0.15 \leq \frac{
 {\rm BR}(B_u\rightarrow\tau \nu_{\tau})_{\rm MSSM}}
 {{\rm BR}(B_u\rightarrow \tau \nu_{\tau})_{\rm SM}}
        \leq 2.41 \; (3\sigma)~. & \nonumber
\end{eqnarray}
\section{Results}
\label{sec:results}
\subsection{Model I}
\label{sec:model I
}

\begin{figure}[ht!]\hspace*{-1.0cm}
\centering
\includegraphics[scale=0.5]{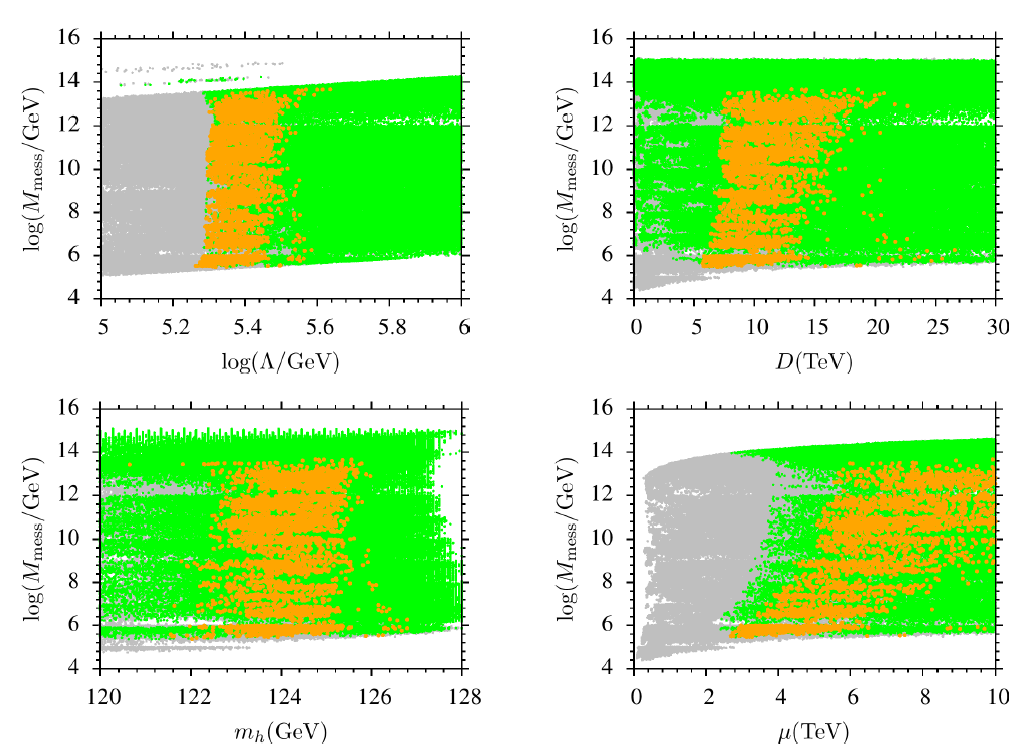}
\caption{Plots in $M_{{\rm mess}}-\Lambda$, $M_{{\rm Mess}}-\mu$, $\Lambda-D$, and $M_{{\rm mess}}-D$ planes. All points are consistent with REWSB. Green points satisfy the mass bounds and constraints from the rare B-meson decays. Yellow points form a subset of green ones, and they represent values of $\Delta a_{\mu}$ that would bring theory and experiment in agreement to within $1\sigma$. }
\label{figure1}
\end{figure}

In this section we present the results of the scan over the parameter space listed in Eq.(\ref{parameterRange}) for Model I. As previously mentioned, the characteristic future of Model I is that all third generation sfermions receive additional universal contribution to their SSB mass compared to the first generation. Figure \ref{figure1} represents plots in $M_{{\rm mess}}-\Lambda$, $M_{{\rm Mess}}-\mu$, $\Lambda-D$, and $M_{{\rm mess}}-D$ planes. All points are consistent with REWSB. Green points satisfy the mass bounds and constraints from rare B-meson decays. Yellow points form a subset of green ones, and they represent values of $\Delta a_{\mu}$ that would bring theory and experiment in agreement to within $1\sigma$. We see from the $M_{{\rm mess}}-\Lambda$ panel that the resolution of muon $g-2$ problem allows only a relatively narrow range for $\Lambda$, $5.2 \lesssim \log(\Lambda/{\rm GeV}) \lesssim 5.6$, while it is possible to find solutions for $5 \lesssim \log(M_{{\rm mess}}/{\rm GeV})\lesssim 14$. Similarly the $\mu-$term can lie in a wide range $\sim 2-10$ TeV consistently with all the experimental constraints and compatible with the muon $g-2$.  The contribution from the flavor $U(1)_1$   symmetry, quantified by $D$, is bounded from the muon $g-2$ constraint  (see  yellow points).
As seen from the $M_{{\rm mess}}-D$ panel, in order to stay within a  1$\sigma$ range of muon $g-2$ and have a light CP even Higgs boson mass  around 125 GeV, the additional contribution from flavor $U(1)_1$ satisfies $5$ TeV $<  D < 20 $ TeV.

In the minimal GMSB scenario, at  the messenger scale the bino and right-handed slepton masses are related to each other, namely $M_1^2=\frac{5}{6} m_{\tilde{\ell}_{R}}^2$  \cite{Dimopoulos:1996yq}. Even though all right-handed sleptons are degenerate at the messenger scale, the  lightest stau can be the lightest slepton at low scale due to the Yukawa coupling contribution to the RGE evolution. Also, there is the tri-linear SSB coupling contribution to the stau mass matrix which can make it much lighter. Hence, in GMSB we have either a stau  or neutralino NLSP depending on the initial parameters. However, in our model, since we have additional positive contributions to the third generation sfermion masses the stau becomes heavier and so the
right-handed smuon or selectron can be the NLSP. Note that a right-handed selectron/smuon  NLSP in gauge mediation models was studied in Ref. \cite{Calibbi:2014pza} without addressing the muon $g-2$ anomaly.  Also the sparticle spectrum are very different in comparison since we have a very different gauge sector.

\begin{figure}[]\hspace*{-1.0cm}
\centering
\includegraphics[scale=0.5]{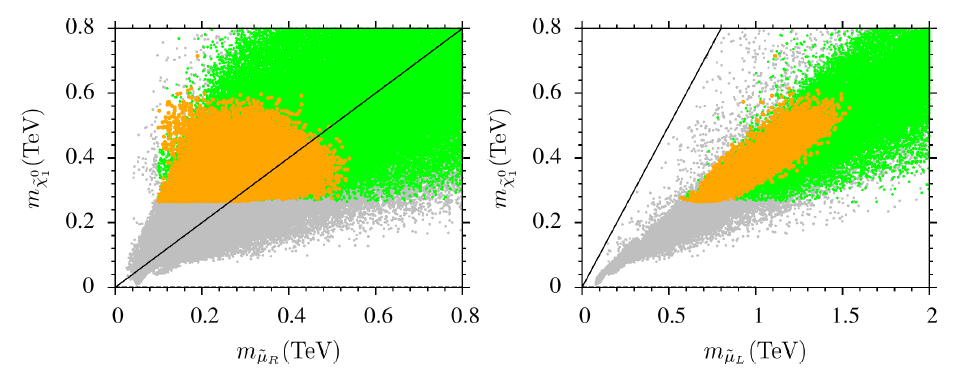}
\caption{Plots in $m_{\tilde{\chi}_{1}^{0}}-m_{\tilde{\mu}_{R}}$, $m_{\tilde{\chi}_{1}^{0}}-m_{\tilde{\mu}_{L}}$ planes. The color coding is the same as in Figure \ref{figure1} except that the mass bound on smuon and neutralino are not applied here. The unit line indicates the regions where the two masses represented on the axes are equal to each other.}
\label{figure2}
\end{figure}

Figure \ref{figure2} displays the results in the $m_{\tilde{\chi}_{1}^{0}}-m_{\tilde{\mu}_{R}}$, $m_{\tilde{\chi}_{1}^{0}}-m_{\tilde{\mu}_{L}}$ planes. The color coding is the same as in Figure \ref{figure1} except that the smuon and neutralino mass bounds are not applied here. The unit line indicates that it is possible to have neutralino - selectron/smuon degeneracy. Note that we have either smuon or neutralino NLSP solutions in the model.
As seen from the $m_{\tilde{\chi}_{1}^{0}}-m_{\tilde{\mu}_{R}}$ panel, there are plenty of solutions for both neutralino NLSP and right-handed smuon NLSP. Even though the neutralino mass bound is not applied in this panel, the lightest neutralino mass cannot be lighter than $250$ GeV due to the current mass bound on gluino ($m_{\tilde{g}} \geq 1.5$ TeV). The reason why the gluino and neutralino masses are tied up is due to the following relation among the gauginos at the messenger scale: $M_1 : M_2 :M _3 = \alpha_1: \alpha_2 : \alpha_3$. Similarly, as mentioned above, the neutralino and the right-handed smuon masses are related to each other ($M_1^2=\frac{5}{6} m_{\tilde{\ell}_{R}}^2$) as well. However, the RGE evolution can lower the right-handed smuon mass to $\sim 100$ GeV, while the neutralino cannot be lighter than 250 GeV. On the other hand, such light smuons can be excluded by the search for prompt decay inside the detector if gravitino is the LSP. In this case, the mass bound on the right-handed smuon NLSP is $m_{\tilde{\mu}}>245$ GeV, and it becomes more severe if the smuon (and selectron) can decay outside the detector ($m_{\tilde{\mu}}>400$ GeV) \cite{Calibbi:2014pza}. All these bounds are applicable if we assume  R-parity conservation. If we assume phenomenologically acceptable R-parity violation in the model, then this bound will disappear and one can have $m_{\tilde{\mu}} \gtrsim 100$ GeV.  Note that the authors in Ref. \cite{Calibbi:2014pza} presented several models and performed the collider phenomenology  with smuon/selectron as NLSP. Our model has a very different sparticle spectrum compared to the models in  Ref. \cite{Calibbi:2014pza} except for the NLSP smuon/selectron.

  The $m_{\tilde{\chi}_{1}^{0}}-m_{\tilde{\mu}_{L}}$ panel in Figure \ref{figure2} also indicates that we have a stringent lower bound for left-handed smuons/selectrons. As for the right-handed smuon and neutralino, the left-handed smuon mass is related to the wino mass as $M_2^2 \simeq \frac{2}{3}m_{\tilde{\ell}_{L}}$ at the messenger scale. Since the gluino mass bound affects the wino mass, it also affects the left-handed smuon mass. This can be seen from the $m_{\tilde{\chi}_{1}^{0}}-m_{\tilde{\mu}_{L}}$ panel in which we only have solutions with $m_{\tilde{\mu}_{L}}\gtrsim 500$ GeV.

\begin{figure}[ht]
\centering
\includegraphics[scale=0.45]{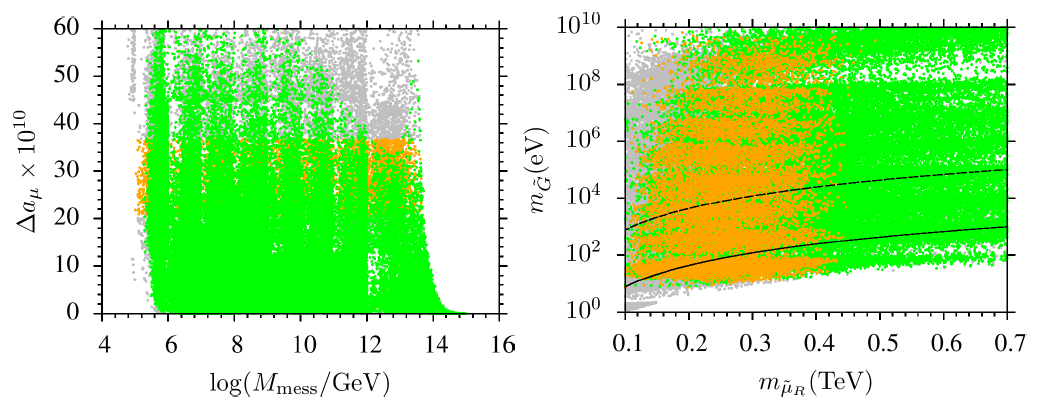}
\caption{Plots in the    $\Delta a_{\mu} - M_{{\rm mess}}  $ and  $m_{\tilde{G}} - m_{\tilde{\mu}_{R}}$ planes. In contrary to other figures the gray points  are consistent with REWSB with the selectron/smuon as NLSP. Green points satisfy the mass bounds and constraints from the rare B-meson decays. Yellow points form a subset of green ones, and they represent values of $\Delta a_{\mu}$ that would bring theory and experiment to within $1\sigma$.  }
\label{figgg-3}
\end{figure}

\begin{figure}[ht!]\hspace*{-1.0cm}
\centering
\includegraphics[scale=0.5]{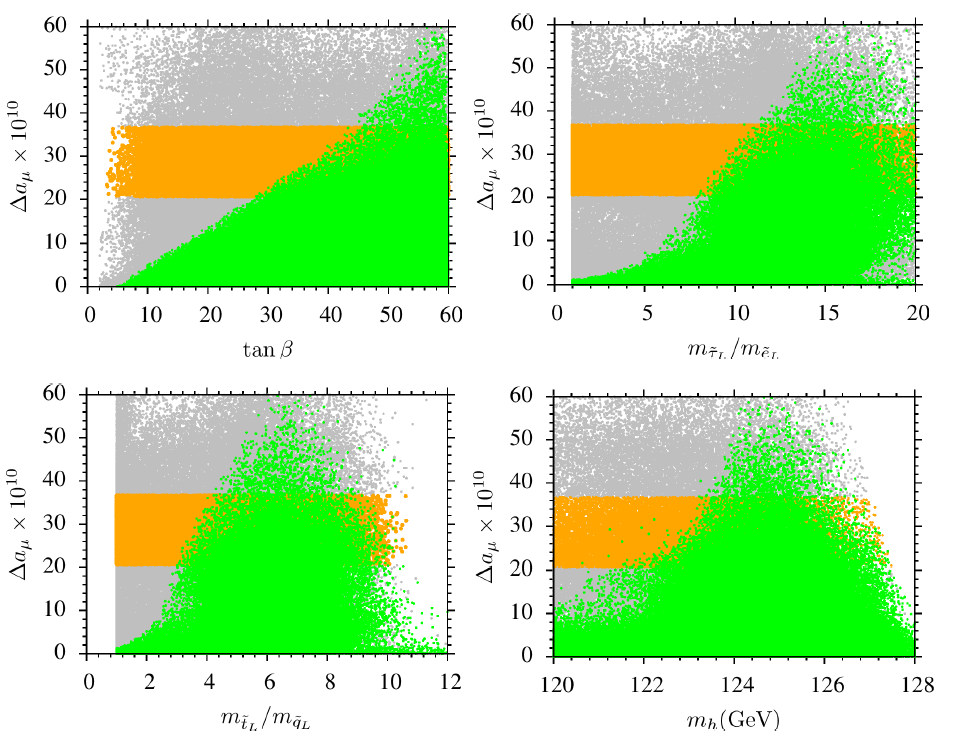}
\caption{Plots in $\Delta a_{\mu}-\tan\beta$, $\Delta a_{\mu}-m_{\tilde{\tau}_{L}}/m_{\tilde{e}_{L}}$, $\Delta a_{\mu}-m_{\tilde{t}_{L}}/m_{\tilde{q}_{L}}$, and $\Delta a_{\mu}-m_{h}$ planes. The color coding is the same as in Figure \ref{figure1} except that yellow points form an independent subset of gray, and they represent the values of $\Delta a_{\mu}$ which would bring theory and experiment to within $1\sigma$. Besides, the Higgs boson mass bound is not applied in the $\Delta a_{\mu}-m_{h}$ plane.}
\label{figure3}
\end{figure}

\begin{figure}[ht!]\hspace*{-1.0cm}
\centering
\includegraphics[scale=0.5]{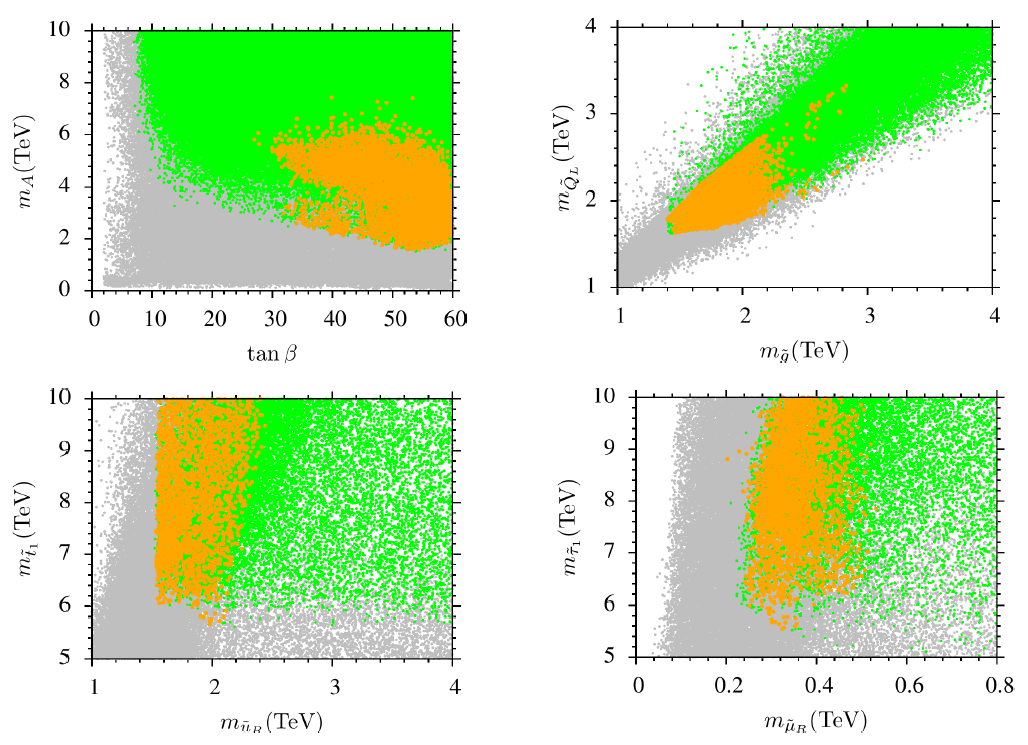}
\caption{Plots in $m_{A}-\tan\beta$, $m_{\tilde{Q}_{L}}-m_{\tilde{g}}$, $m_{\tilde{t}_{1}}-m_{\tilde{u}_{R}}$, and $m_{\tilde{\tau}_{1}}-m_{\tilde{\mu}_{R}}$ planes. The color coding is the same as in Figure \ref{figure1}.}
\label{figure4}
\end{figure}

In Figure \ref{figgg-3} we illustrate our results in the    $\Delta a_{\mu} - M_{{\rm mess}}  $ and  $m_{\tilde{G}} - m_{\tilde{\mu}_{R}}$ planes. Contrary to previous figures, the gray points  are consistent with REWSB and represent solutions with a selectron/smuon NLSP. Green points satisfy the mass bounds and constraints from the rare B-meson decays. Yellow points form a subset of green ones, and they represent values of $\Delta a_{\mu}$ that would bring theory and experiment in agreement to within $1\sigma$.  The  $\Delta a_{\mu} - M_{{\rm mess}}$ plane shows that  we have the desired contribution to the muon $g-2$ calculation from supersymmetric sparticles for almost any value  of $M_{{\rm mess}}$ with smuon NLSP. On the other hand, this broad range  of $M_{{\rm mess}}$ provides a very  wide interval for the gravitino mass which, in gauge mediation, can be expressed as
%
\begin{figure}[ht]\hspace{-1.0cm}
\includegraphics[scale=0.45]{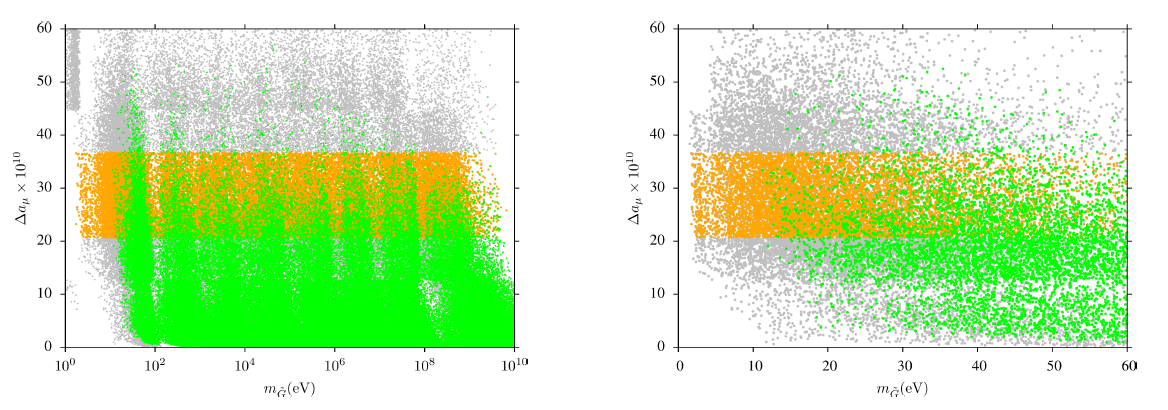}
\caption{Plot in $\Delta a_{\mu}-m_{\tilde{G}}$ plane. The color coding is the same as in Figure \ref{figure3}.}
\label{figure5}
\end{figure}

\begin{eqnarray}
m_{\tilde{G}}\simeq \frac{M_{{\rm mess}}\, \Lambda}{\sqrt{3}\,M_{Pl}}.
\label{grav33}
\end{eqnarray}
where $M_{Pl}$ is the Planck mass.  As shown  in the $M_{{\rm mess}}-\Lambda$ plane of Figure \ref{figure1}, the solutions corresponding to the experimental measurement  for muon $g-2$  within 1$\sigma$ deviation are located in the very narrow interval $1.2 \times 10^{5}\,{\rm GeV}< \Lambda < 4 \times 10^{5}$GeV. Having solutions in a such narrow interval for $\Lambda$ makes the gravitino mass almost linearly dependent on the scale $M_{{\rm mess}}$.

In the second plot in Figure \ref{figgg-3} we show the relation between  $m_{\tilde{G}}$ and $m_{\tilde{\mu}_{R}}$ with smuon being the NLSP. This relationship is very important, as mentioned above, in estimating the smuon life-time which decays into a lepton and a gravitino. In general, for the  NLSP slepton ($\widetilde{\ell}_R$) case the two-body decay width is given by \cite{Martin:1997ns}:
 \begin{eqnarray}
\Gamma ( \widetilde{\ell}_R \to \ell \tilde G )=\frac{m_{\widetilde{\ell}_R}^5}{16 \pi (M_{\rm mess} \Lambda)^2} \left(1-\frac{m_{\ell}^2}{m_{\widetilde{\ell}_R}^2}  \right)^4.
\label{decay33}
\end{eqnarray}
 Using Eq.(\ref{decay33}) and following the discussion in Ref.  \cite{Calibbi:2014pza} we show our results in  the $m_{\tilde{G}} - m_{\tilde{\mu}_{R}}$ plane with a solid and a dashed line. The solid line corresponds to 1 mm free length of the smuon, while the dashed line to 10 m which is supposedly the length of the detector. If a prompt decay is identified with the 1 mm path in which the smuons fly before decaying to a gravitino and a lepton, the region between the solid and dashed lines can be identified as smuon decaying inside the detector. According to the results  represented in the $m_{\tilde{G}} - m_{\tilde{\mu}_{R}}$ plane, Model I predicts a smuon NLSP of mass upto 400 GeV. Hence, the model predicts plenty of solutions testable inside the detectors, even if one assumes R-parity conservation which excludes the region with $m_{\tilde{\mu}_{R}}\lesssim 245$ GeV.

In Figure \ref{figure3} we show  the results  in $\Delta a_{\mu}-\tan\beta$, $\Delta a_{\mu}-m_{\tilde{\tau}_{L}}/m_{\tilde{e}_{L}}$, $\Delta a_{\mu}-m_{\tilde{t}_{L}}/m_{\tilde{q}_{L}}$, and $\Delta a_{\mu}-m_{h}$  planes. The color coding is the same as in Figure \ref{figure1} except that yellow points form an independent subset of gray, and they represents the values of $\Delta a_{\mu}$ which would bring theory and experiment into agreement within $1\sigma$. Besides, the Higgs boson mass bound is not applied in the $\Delta a_{\mu}-m_{h}$ panel. The $\Delta a_{\mu}-\tan\beta$ plane shows the $\tan\beta$ dependence of muon $g-2$. Since the contribution from the supersymmetric particles to the muon $g-2$ depends linearly on $\tan\beta$ it is understandable why there are more solutions satisfying the muon $g-2 $ constraint with increasing $\tan\beta$.  According to the results in this plane, it is hard to have substantial contribution to muon $g-2$ if $\tan\beta \lesssim 30$.

The $\Delta a_{\mu}-m_{\tilde{\tau}_{L}}/m_{\tilde{e}_{L}}$ and $\Delta a_{\mu}-m_{\tilde{t}_{L}}/m_{\tilde{q}_{L}}$ planes highlight the mass splitting necessary between the third and first two-family sfermion SSB masses in order to satisfy all current phenomenological constraints including muon $g-2$. As seen from the $\Delta a_{\mu}-m_{\tilde{\tau}_{L}}/m_{\tilde{e}_{L}}$ plane, the muon $g-2$ solution corresponds to the slepton splitting requirement $m_{\tilde{\tau}_{L}}/m_{\tilde{e}_{L}} \gtrsim 8$ at low scale.  At the same time for the squarks we have a relatively milder constraint as $3 \lesssim m_{\tilde{t}_{L}}/m_{\tilde{q}_{L}} \lesssim 10$. Finally, the $\Delta a_{\mu}-m_{h}$ plane shows plenty of solutions which simultaneously accommodate the 125 GeV Higgs boson and muon $g-2$ anomaly.

We summarize the sparticle spectrum in Figure \ref{figure4} with plots in $m_{A}-\tan\beta$, $m_{\tilde{Q}_{L}}-m_{\tilde{g}}$, $m_{\tilde{t}_{1}}-m_{\tilde{u}_{R}}$, and $m_{\tilde{\tau}_{1}}-m_{\tilde{\mu}_{R}}$ planes. The color coding is the same as in Figure \ref{figure1}. The $m_{A}-\tan\beta$ plane shows that $m_{A}$ can be as low as 2 TeV for $\tan\beta \gtrsim 30$, which is required also by muon $g-2$.
It is very interesting to note that  the solutions which
accommodate the  muon $g-2$ anomaly at 1$\sigma$ level predict squarks and gluino masses lighter than 3.5 TeV. Furthermore, as seen from  the $m_{\tilde{Q}_{L}}-m_{\tilde{g}}$ plane most of the solutions are associated with the gluino masses less than 3 TeV. This makes our model readily testable at LHC Run II. The lightest stop is found to be relatively heavy in order to accommodate the 125 GeV Higgs boson in the model. According to the $m_{\tilde{t}_{1}}-m_{\tilde{u}_{R}}$ panel, $m_{\tilde{t}_{1}} \gtrsim 5.5$ TeV for $m_{\tilde{u}_{R}} \gtrsim 2$ TeV. Finally, as seen from the $m_{\tilde{\tau}_{1}}-m_{\tilde{\mu}_{R}}$, $m_{\tilde{\mu}_{R}}$ plane is bounded in the $\sim 200-500$ GeV range as expected, since the supersymmetric contributions to muon $g-2$ rely on the light smuons. Again, $m_{\tilde{\tau}_{1}} \gtrsim 5.5$ TeV because of the flavor symmetry contribution.

We present our results for the gravitino mass in Figure \ref{figure5} with the $\Delta a_{\mu}-m_{\tilde{G}}$ panel. The color coding is the same as in Figure \ref{figure3}. In GMSB scenarios the gravitino is usually the LSP, and its mass can vary between 10 TeV and a few eV. As seen from the left panel of Figure \ref{figure5}, it is possible to stay within the $1\sigma$ band of muon $g-2$ for a wide range of the gravitino mass, from a few eV to 10 GeV. A light gravitino provides a plausible dark matter candidate and it can also manifest itself through missing energy in colliders \cite{Viel:2005qj}. In standard scenarios, the relic density bound ($\Omega h^{2}\simeq 0.11$ \cite{Hinshaw:2012aka}) is satisfied with a gravitino mass $\sim 200$ eV \cite{Viel:2005qj}, which makes the gravitino a hot dark matter candidate. Hot dark matter, however, cannot compose more than $15\%$ of the total dark matter density and this, in turn, implies that the gravitino mas should be less than 30 eV \cite{Viel:2005qj}. The right panel of Figure \ref{figure5} is the same as the left panel, except that it is zoomed to the region which yields gravitino masses of the order of a few eV. The predicted gravitino mass compatible with muon $g-2$ is consistent with the hot dark matter gravitino, since  resolution of muon $g-2$ within the sGMSB framework requires $m_{\tilde{G}} \gtrsim 10$ eV as seen from the right panel. In order to have a complete dark matter scenario one could invoke axions as cold dark matter in this region.

A gravitino mass $\gtrsim$ 30 eV requires non-standard scenarios  in order to agree with observations.
 Such non-standard scenarios include gravitino decoupling and freezing out earlier than in the standard scenario, which may be possible in a theory with more degrees of freedom than the MSSM \cite{Feng:2010ij}.   A gravitino of mass $\gtrsim {\rm \ keV}$ is still possible  and it can be cold enough to constitute all of the dark matter if non-standard scenarios such as early decoupling is assumed.

\begin{table}[tp!]
\centering
\scalebox{0.89}{
\begin{tabular}{|l|ccc|}
\hline
                 & Point 1 & Point 2 & Point 3  \\
\hline
\hline
$\Lambda$       & $0.25\times 10^{6}$  & $0.26\times 10^{6}$ & $0.34\times 10^{6}$ \\
$M_{\rm mess}$ & $0.9\times 10^{6}$ & $0.50\times 10^{6}$ & $0.20\times 10^{7}$ \\
$\tan\beta$    & 58 & 57 & 59 \\
$D$            & 11580 & 9755 & 22060 \\
$m_{\tilde{\tau}_{1}}/m_{\tilde{e}_{L}}$ & 14 & 11 & 19 \\
$m_{\tilde{t}_{1}}/m_{\tilde{u}_{L}}$ & 5 & 4 & 7 \\
\hline
&&& \\
$\Delta a_{\mu}$ & $28.9\times 10^{-10}$  & $21\times 10^{-10}$ & $23\times 10^{-10}$ \\
&&& \\
\hline
$\mu$          & 5404 & 4424 & 10179 \\
$A_{t}$        & -374 & -371 & -496 \\
$A_{b}$        & -385 & -382 & -509 \\
$A_{\tau}$     & -36 & -36 & -52 \\
\hline
$m_h$           & 125 & 125 & 126 \\
$m_H$           & 2201  & 1954  & 3504 \\
$m_A$           & 2187 & 1942 & 3481 \\
$m_{H^{\pm}}$   & 2204 & 1957 & 3505 \\

\hline
$m_{\tilde{\chi}^0_{1,2}}$
                 & \textbf{347}, 678  &  \textbf{371},723  & 483, 936  \\

$m_{\tilde{\chi}^0_{3,4}}$
                 & 5312, 5312  & 4347, 4347  & 10005, 10005 \\

$m_{\tilde{\chi}^{\pm}_{1,2}}$
                & 680, 5245 & 724, 4292 & 939, 9879 \\

$m_{\tilde{g}}$  & 1902 & 2012 & 2563 \\
\hline $m_{ \tilde{u}_{L,R}}$
                 & 2396, 2279  & 2561, 2440  & 3060, 2898 \\
$m_{\tilde{t}_{1,2}}$
                 & 11084, 11332 & 9483, 9770 & 20940, 21360 \\
\hline $m_{ \tilde{d}_{L,R}}$
                 & 2397, 2266  & 2562, 2427  & 3061, 2879 \\
$m_{\tilde{b}_{1,2}}$
                 & 11275, 11423  & 9648, 9770  & 21276, 21528 \\
\hline
$m_{\tilde{\nu}_{e,\mu}}$
                 & 798 & 845 & 1031 \\
$m_{\tilde{\nu}_{\tau}}$
                 & 11407 & 9642  & 21654 \\
\hline
$m_{ \tilde{\mu}_{L,R}}$
                & 853, 363  & 883, 398  & 1170, \textbf{405} \\
$m_{\tilde{\tau}_{1,2}}$
                & 11067, 11399 & 9366, 9637  & 20976, 21360 \\
\hline
$m_{\tilde{G}}$ (eV) & 54 & 31 & 157 \\
\hline
\end{tabular}}
\caption{Benchmark points for exemplifying our results. All points are chosen to be consistent with the experimental constraints and muon $g-2$. All masses and scales are given in GeV units except for the gravitino mass which is given in eV. Point 1 represents a solution with a 125 GeV Higgs boson, $\Delta a_{\mu}\approx 28.7\times 10^{-10}$ with the lowest possible messenger scale ($M_{{\rm }} \simeq 10^{6}$). Point 2 depicts a solution with the least amount of mass splitting within the squark and slepton families compatible with the 125 GeV Higgs and desired muon $g-2$ ($m_{\tilde{\tau}_{1}}/m_{\tilde{e}_{L}} \approx 11$, $m_{\tilde{t}_{1}}/m_{\tilde{u}_{L}} \approx 4$). Points 1 and 2 display solutions with neutralino NLSP, while point 3 shows a solution with smuon NLSP which is compatible with the Higgs boson mass, muon $g-2$.}
\label{benchsgmsb}
\end{table}

Finally we present three benchmark points in Table \ref{benchsgmsb} which exemplify our findings. All points are chosen to be consistent with the experimental constraints and muon $g-2$. All masses and scales are given in GeV units except the gravitino mass which is given in eV. Point 1 represents a solution with a the 125 GeV Higgs boson and $\Delta a_{\mu}\approx 28.7\times 10^{-10}$ with the lowest possible messenger scale ($M_{{\rm }} \simeq 10^{6}$). Point 2 depicts a solution with the least amount of mass splitting within the squark and slepton families compatible with the 125 GeV Higgs and desired muon $g-2$ ($m_{\tilde{\tau}_{1}}/m_{\tilde{e}_{L}} \approx 11$, $m_{\tilde{t}_{1}}/m_{\tilde{u}_{L}} \approx 4$). Points 1 and 2 display solutions with neutralino NLSP, while point 3 shows a solution for the smuon NLSP which is compatible with the Higgs boson mass, muon $g-2$.

\subsection{Model II}
\label{modelII}
In this section we present our results for a different charge configuration under the extra $U(1)$ symmetry listed as Model II in Table \ref{table 1}.  Note that the fundamental parameters remain the same, and only the contribution from the extra $U(1)_{2}$ symmetry to the sparticles is now different. In Model II, the third family squark masses receive  additional contributions, as in Model I, parameterized by $D^{2}$.  On the other hand  the third family charged slepton mass$^2$ receives the contribution of $9D^{2}$. As we will show below, having stau heavier than stop has  consequences for muon $g-2$ contribution and the Higgs boson mass.


\begin{figure}[!htp]
\centering
\includegraphics[scale=1.1]{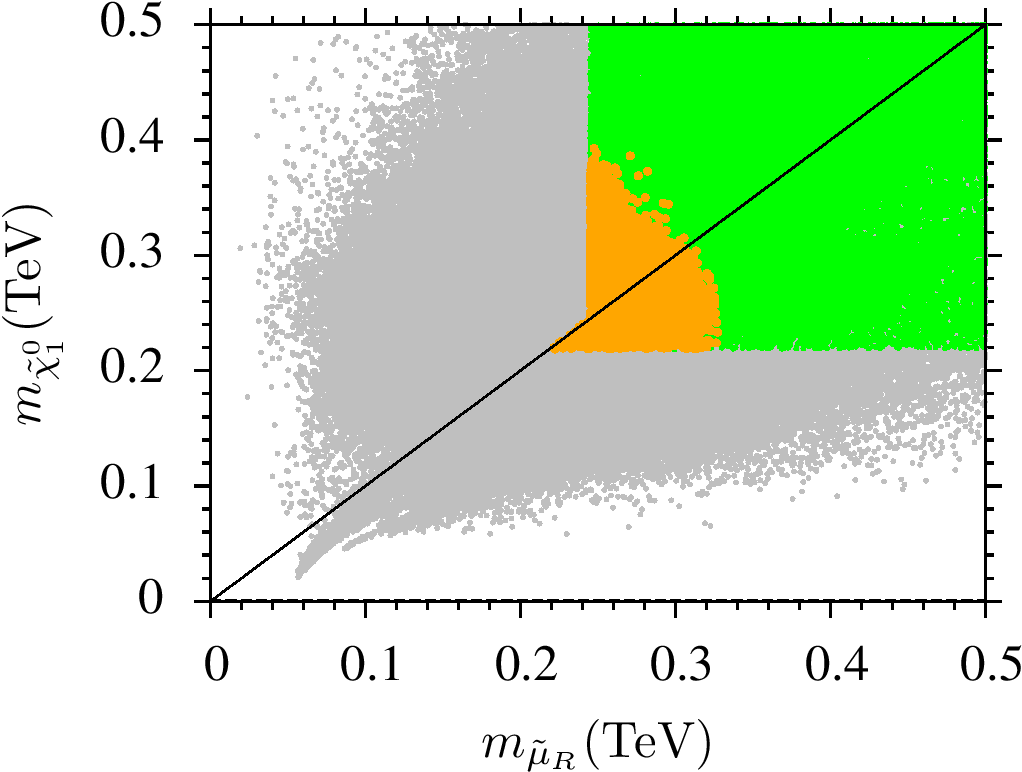}
\caption{Plot in the $m_{\tilde{\chi}_{1}^{0}}-m_{\tilde{\mu}_{R}}$ plane. The color coding is the same as in Figure \ref{figure1}.}
\label{fig-1-bl}
\end{figure}


\begin{figure}[h]\hspace{-1.0cm}
\centering
\includegraphics[scale=0.45]{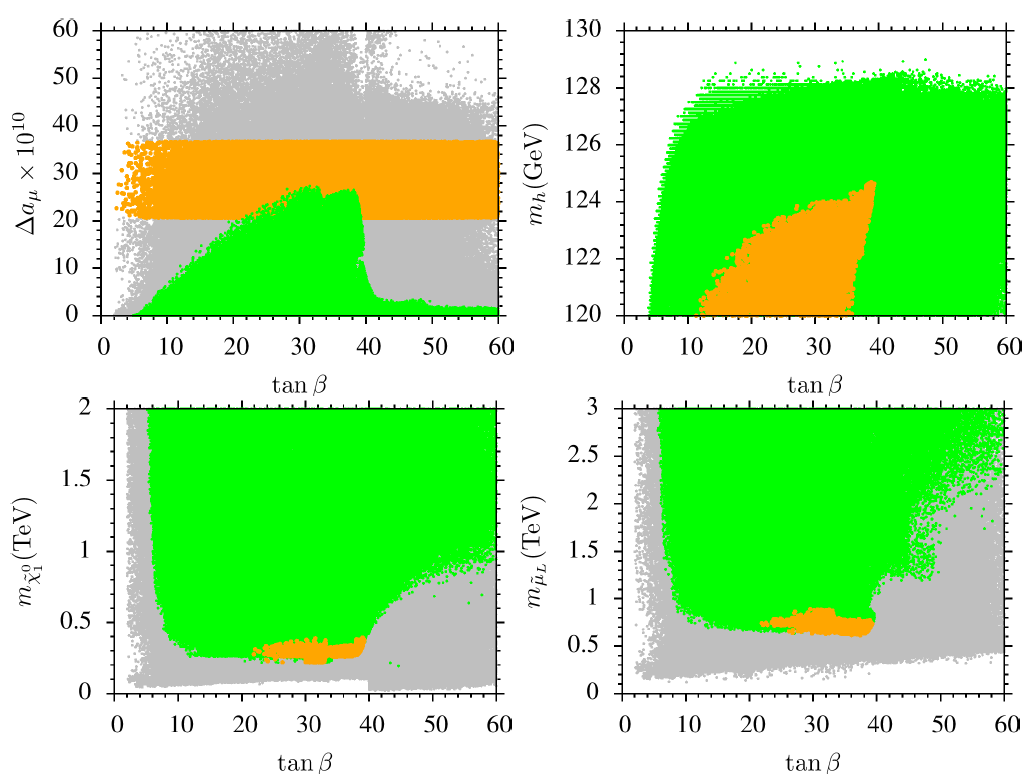}
\caption{Plots in the $\Delta a_{\mu}-\tan\beta$, $m_{h}-\tan\beta$, $m_{\tilde{\chi}_{1}^{0}}$, and $m_{\tilde{\mu}_{L}}-\tan\beta$ planes. The color coding is the same as in Figure \ref{figure1} except the $\Delta a_{\mu}-\tan\beta$ plane whose color coding is the same as in Figure \ref{figure4}. }
\label{fig-2bl}
\end{figure}
\begin{figure}[]\hspace{-1.0cm}
\centering
\includegraphics[scale=0.45]{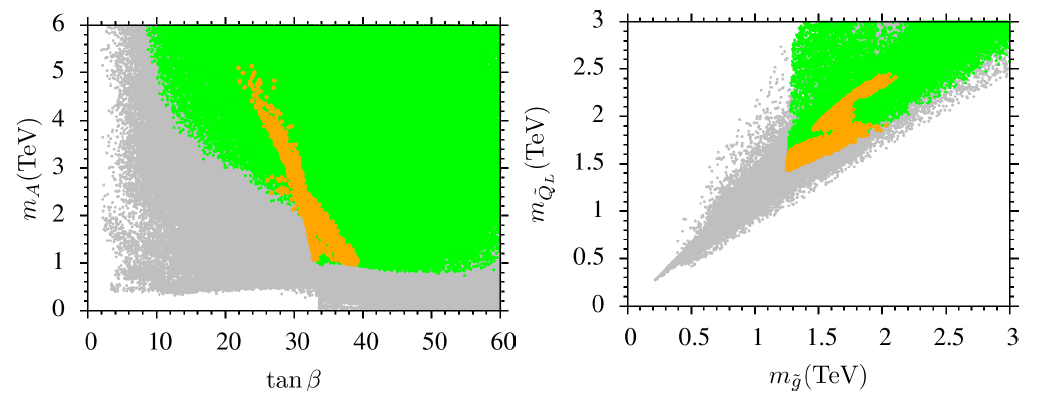}
\caption{Plots in the $m_{A}-\tan\beta$ and $m_{\tilde{Q}_{L}}-m_{\tilde{g}}$ planes. The color coding is the same as in Figure \ref{figure1}.}
\label{fig-3bl}
\end{figure}


\begin{figure}[]
\centering
\hspace{-1.0cm}\includegraphics[scale=0.7]{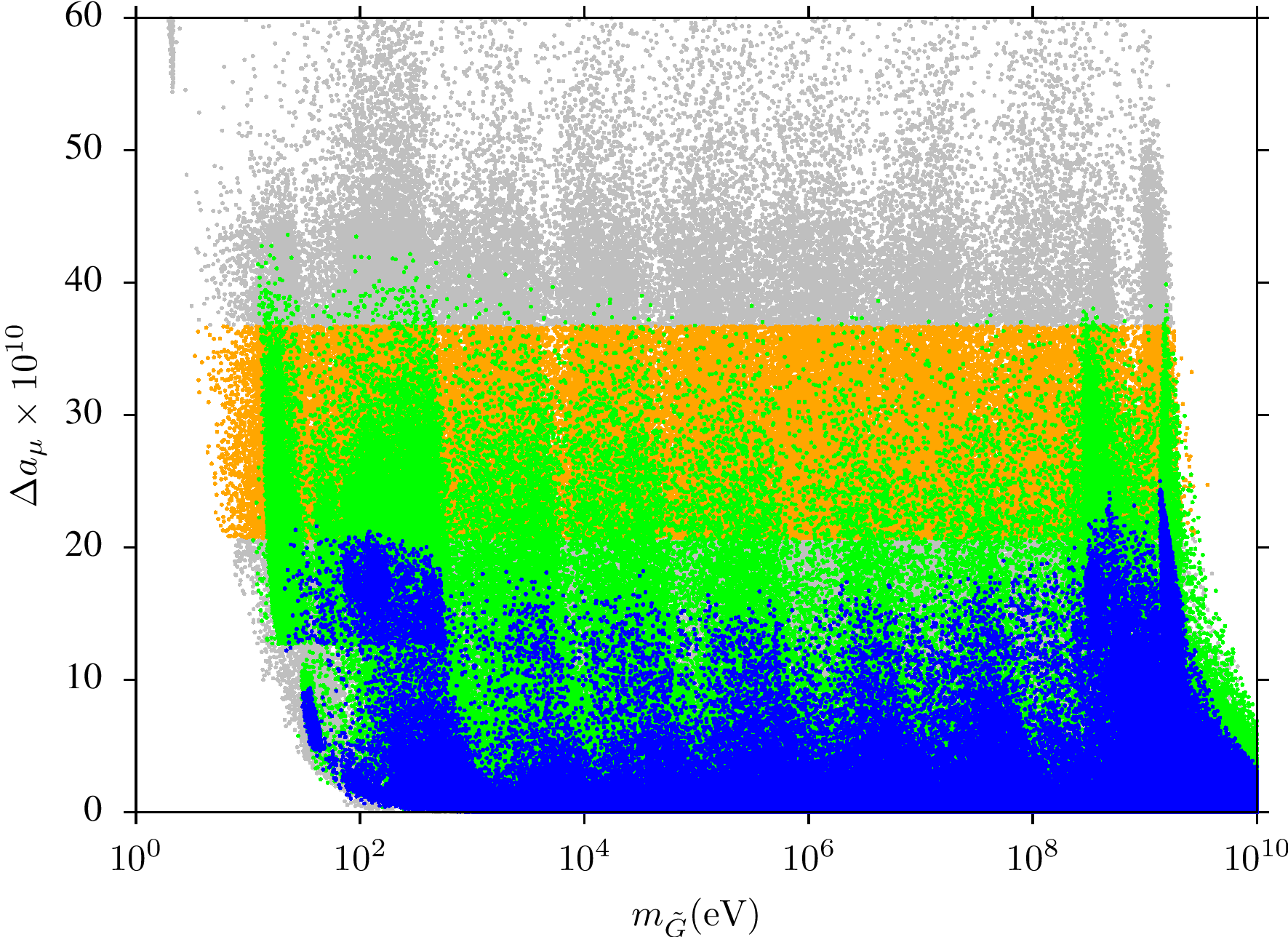}
\caption{Plots in the $\Delta a_{\mu}-m_{\tilde{G}}$ and $m_{\tilde{G}}-m_{\tilde{\mu}_{R}}$ planes. All points are consistent with REWSB. The color coding is the same as in Figure \ref{figure4}. In addition, the blue points represent the solutions with neutralino NLSP. The $m_{\tilde{G}}-m_{\tilde{\mu}_{R}}$ plane plots only the solutions with smuon NLSP. The color coding used in this plane is same as in Figure \ref{figure1}.}
\label{fig-4bl}
\end{figure}

 We present our results  for Model II first in terms of the parameters $m_{\tilde{\chi}_{1}^{0}}$ and $m_{\tilde{\mu}_{R}}$, which are some of the crucial players for muon $g-2$ anomaly.
 The  $m_{\tilde{\chi}_{1}^{0}}-m_{\tilde{\mu}_{R}}$ plot is shown in Figure  \ref{fig-1-bl} with the color coding the same as in Figure \ref{figure1}. The results displayed in the $m_{\tilde{\chi}_{1}^{0}}-m_{\tilde{\mu}_{R}}$ plane exhibit sharp lower bounds on the neutralino and right-handed smuon masses. As in Model I, gluino mass bound determines the lower bound on the neutralino mass. We have two different bounds for  the right-handed smuon mass. If the lightest smuon is heavier than the neutralino, the lower mass bound is 100 GeV \cite{pdg}. When smuon is NLSP it decays to gravitino and muon and the lower mass bound 245 GeV. This bound can be go up to 400 GeV when  gravitino is heavier then keV or so. But this bound can be relaxed up to 245 GeV if we assume suitable R-parity violation in the theory.
 The tail shape in the plot corresponding to the region with $m_{\tilde{\mu}_{R}} \gtrsim m_{\tilde{\chi}_{1}^{0}}$, and since the right-handed smuon is not NLSP in this region, the $\tilde{\mu}\rightarrow \mu \tilde G$ decay constraint is not applicable. As we can see  in this model neutralino  can be as light as about 200 GeV or so. There is difference compare to the Model I in terms of upper bound for neutralino and smuon. Here the neutralino needs to be lighter than 400 GeV and smuon lighter than 320 GeV in order to bring the muon $g-2$ into agreement within 1$\sigma$ of the experimental results. It was noticed in Ref. \cite{Ajaib:2015yma} that a smuon in this mass interval can be tested at the LHC.  On the other hand, the bound for this particle in model I was up to 600 GeV. The reduction of the upper bound of this particle can be related to the $\tan\beta$ bounds for model II which we display below.

We use Figure  \ref{fig-2bl} to illustrate the impact of the extra heavy stau lepton in the spectrum and the results are presented in the $\Delta a_{\mu}-\tan\beta$, $m_{h}-\tan\beta$, $m_{\tilde{\chi}_{1}^{0}}-\tan\beta$, and $m_{\tilde{\mu}_{L}}-\tan\beta$ planes. The color coding is the same as in Figure \ref{figure1} except for the $\Delta a_{\mu}-\tan\beta$ plane where the color coding is the same as in Figure \ref{figure4}. The $\Delta a_{\mu}-\tan\beta$ plane reveals a sharp fall in muon $g-2$ values (green points) for $\tan\beta \approx 40$. One would have expected the supersymmetric contributions to muon $g-2$ to linearly increase with $\tan\beta$ \cite{Moroi:1995yh}  as we found for Model I. This sharp fall can be understood in terms of the conflict between the light CP-even Higgs boson mass and muon $g-2$ calculation. We see from the $\Delta a_{\mu}-\tan\beta$ plane that there are plenty of solutions (green points) that satisfy all collider constraints and the requirement that $124 \,{\rm GeV} \leq m_h \leq 127 \, {\rm GeV}$ with  $\tan\beta\geq40$.  However, all of these solution provide negligible contribution to the muon $g-2$ calculation. We conclude that for $\tan\beta\geq40$ we have heavy smuons or neutralinos in the spectrum in order to obtain the required  Higgs boson mass.
 This observation can be related to the results presented in Ref. \cite{Brignole:2002bz}, namely that the bottom squark and stau can provide negative contribution to the light CP even Higgs boson mass calculation. This contribution can become  significant for $\tan\beta \geq 40$. In order to compensate this and obtain correct Higgs boson mass, we need to increase the contribution from the stop loop. This, in turn, trough RGE evolution increases the gaugino and  smuon masses. The $m_{h}-\tan\beta$ plane confirms this conclusion. Here we can see that it is possible to obtain the Higgs boson mass compatible with the experimental data (green points), with the muon $g-2$ constraint presented in yellow points.

Exploring the  $m_{\tilde{\chi}_{1}^{0}}-\tan\beta$ and $m_{\tilde{\mu}_{L}}-\tan\beta$ planes we found that the arc shape (green points) in these plots is caused by the Higgs boson mass requirement. Here we can see that the neutralino and smuon masses drastically increase for  $\tan\beta \geq 40$, which leads to a correponding decrease in the supersymmetric contribution to the muon $g-2$ calculation.

Figure \ref{fig-3bl} represents the sparticle spectrum in the $m_{A}-\tan\beta$ and $m_{\tilde{Q}_{L}}-m_{\tilde{g}}$ planes. The color coding is the same as in Figure \ref{figure1}. The $m_{A}-\tan\beta$ plane shows that $m_{A}$ can be as light as 1 TeV or so. This bound is also compatible with the muon $g-2$ condition. Similarly the $m_{\tilde{Q}_{L}}-m_{\tilde{g}}$ plane reveals relatively light squark and gluino masses. According to our results, muon $g-2$ requires $m_{\tilde{Q}_{L}} \lesssim 2.5$ TeV, while $m_{\tilde{g}}\lesssim 2$ TeV. In contrast to the usual GMSB framework, the first two family squarks are lighter than the third family and quite possibly accessible at the LHC 13.

Figure \ref{fig-4bl}  displays our results in the $\Delta a_{\mu}-m_{\tilde{G}}$ plane, The color coding is the same as in Figure \ref{figure4} and all points are consistent with REWSB. The blue points represent solutions with neutralino NLSP.  As seen from the $\Delta a_{\mu}-m_{\tilde{G}}$ plane, the gravitino mass can lie in a wide range from a few eV to 10 GeV, consistent with the muon $g-2$ constraint. The blue region indicates solutions with neutralino NLSP which are not constrained by the prompt  smuon decay. The residual green region is the set of solutions with smuon NLSP. The results for the gravitino mass are similar to those obtained in Model I that a gravitino mass lighter then 30 eV is only realized if the smuon is NLSP.  However, in Model II, the solutions with neutralino NLSP , that are consistent with muon $g-2$ anomaly yield a gravitino with mass of the order of GeV, which makes it difficult to have the latter as dark matter particle.

Finally, we present two benchmark points in Table \ref{benchsgmsb2} to exemplify our findings for Model II . All points are chosen to be consistent with the experimental constraints and muon $g-2$. All masses and scales are given in GeV units except for the gravitino mass which is given in eV. Point 1 displays a solution with the largest $\tan\beta$ value compatible with muon $g-2$. The NLSP happens to be the right-handed smuon for Point 1.The staus are heavy, which push the right-handed smuon masses to low values through RGEs (in contrast to the left-handed smuons). Hence the large $\tan\beta$ region mostly yields right-handed smuon NLSP as displayed in the plots above. In addition, Point 1 represents a solution with the gravitino mass about 19 eV. Similarly, Point 2 represents a solution with gravitino mass of about 30 eV. The spectrum is quite similar to Point 1, and one can compare how a slight change in the stau mass can affect the right-handed smuon mass. The stau mass is slightly lighter in Point 2, which leads to a slightly heavier right-handed smuon with mass about 255 GeV in the spectrum. The benchmark points in Table \ref{benchsgmsb2} exemplify the impact of heavy staus in the low energy phenomenology.

\begin{table}[]
\centering
\scalebox{0.89}{
\begin{tabular}{|l|cc|}
\hline
                 & Point 1 & Point 2   \\
\hline
\hline
$\Lambda$       & $0.24\times 10^{6}$  & $0.24\times 10^{6}$  \\
$M_{\rm mess}$ & $0.32\times 10^{6}$ & $0.57\times 10^{6}$  \\
$\tan\beta$    & 39 & 38 \\
$D$            & 11100 & 10060  \\
$m_{\tilde{\tau}_{1}}/m_{\tilde{e}_{L}}$ & 13.7 & 12.5 \\
$m_{\tilde{t}_{1}}/m_{\tilde{u}_{L}}$ & 4.9 & 4.6  \\
\hline
&& \\
$\Delta a_{\mu}$ & $20.8\times 10^{-10}$  & $21.0\times 10^{-10}$  \\
&& \\
\hline
$\mu$          & 4437 & 4347  \\
$A_{t}$        & -334 & -350  \\
$A_{b}$        & -351 & -371  \\
$A_{\tau}$     & -39 & -41  \\
\hline
$m_h$           & 125 & 124  \\
$m_H$           & 1143  & 1260  \\
$m_A$           & 1136 & 1252  \\
$m_{H^{\pm}}$   & 1146 & 1263  \\

\hline
$m_{\tilde{\chi}^0_{1,2}}$
                 & 382, 743  & 341, 667   \\

$m_{\tilde{\chi}^0_{3,4}}$
                 & 4364, 4364  & 4277, 4277  \\

$m_{\tilde{\chi}^{\pm}_{1,2}}$
                & 744, 4304 & 667, 4221  \\

$m_{\tilde{g}}$  & 2058 & 1868  \\
\hline $m_{ \tilde{u}_{L,R}}$
                 & 2441, 2338  & 2350, 2246  \\
$m_{\tilde{t}_{1,2}}$
                 & 10785, 11126 & 9705, 10066  \\
\hline $m_{ \tilde{d}_{L,R}}$
                 & 2443, 2339  & 2352, 2246  \\
$m_{\tilde{b}_{1,2}}$
                 & 11085, 11281  & 10025, 10232  \\
\hline
$m_{\tilde{\nu}_{e,\mu}}$
                 & 688  & 686  \\
$m_{\tilde{\nu}_{\tau}}$
                 & 33152 & 30021  \\
\hline
$m_{ \tilde{\mu}_{L,R}}$
                & 770, \textbf{246}  & 763, \textbf{255}  \\
$m_{\tilde{\tau}_{1,2}}$
                & 32938, 33182  & 29797, 30046  \\
\hline
$m_{\tilde{G}}$ (eV) & 19 & 33 \\
\hline
\end{tabular}}
\caption{Benchmark points for Model II. All points are chosen to be consistent with the experimental constraints and muon $g-2$. All masses and scales are given in GeV units except for the gravitino mass which is given in eV. Point 1 displays a solution with the largest $\tan\beta$ value compatible with muon $g-2$. The NLSP happens to be the right-handed smuon for Point 1. Point 2 depicts a solution with neutralino NLSP. }
\label{benchsgmsb2}
\end{table}

\section{Conclusion}
\label{sec:conclusion}
We have explored the sparticle and Higgs phenomenology of a flavor symmetry based MSSM  in the framework of gauge mediation supersymmetry breaking scenario.  Explicit ultra-violet completion of models that generate a symmetry based MSSM  (sMSSM) spectrum  \cite{Babu:2014sga}   at low energies have been presented. These include models based on an additional U(1) flavor symmetry which acts only on the  third generation fermions. The charge assignments of these models (Type I \& 2) are given in Table 1. Our model  contains one  additional parameter compared to the canonical GMSB scenario. Specifically we have a common SSB mass term for the first two family sfermions (which are lighter  in mass compared to the third family sfermions.). This splitting between the family masses is generated by exploiting the U(1)symmetry which only acts on the third generation. This  helps us to reconcile the muon $g-2$ anomaly with the Higgs boson mass, and we can also accommodate the desired relic abundance of dark matter in the form of gravitino.  We find that the simultaneous explanation of these observables  requires the gluino and first two family squark masses to be less than 2.5 TeV. The mass of the first  two family sleptons is less than 600 GeV (or so) in Model I and less than 300 GeV for Model II.


\section*{Acknowledgments}

 We thank Howard  Baer and Azar Mustafayev for helpful discussions.  This work is supported in part by the DOE Grant DE-SC0013880 (I.G. and Q.S.) and The Scientific and Technological Research Council of Turkey (TUBITAK) Grant no. MFAG-114F461 (CS\"{U}). This work used the Extreme Science and Engineering Discovery Environment (XSEDE), which is supported by the National Science Foundation grant number OCI-1053575.  I.G. acknowledges support from the  Rustaveli National Science Foundation  No. 03/79. Part of the numerical calculations reported in this paper were performed at the National Academic Network and Information Center (ULAKBIM) of TUBITAK, High Performance and Grid Computing Center (TRUBA Resources).

\end{document}